\documentclass[aps,twocolumn,showpacs,superscriptaddress]{revtex4}
\usepackage{epsfig}

\def\avg#1{\langle#1\rangle}

\def\be{\begin{equation}}       \def\ee{\end{equation}}
\def\bea{\begin{eqnarray}}      \def\eea{\end{eqnarray}}
\def\PRB{Phys. Rev. B}
\def\PRL{Phys. Rev. Lett.}

\begin{document}

\title{Vortex configurations of bosons in an optical lattice
}
\author{Congjun Wu}
\affiliation{Department of Physics, McCullough Building, Stanford
University, Stanford CA 94305-4045}
\author{Han-dong Chen}
\affiliation{Department of Applied Physics, McCullough Building,
Stanford University, Stanford CA 94305-4045}
\author{Jiang-ping Hu}
\affiliation{Department of Physics and Astronomy, University of
California, Los Angels, CA 90095-1547}
\author{Shou-Cheng Zhang}
\affiliation{Department of Physics, McCullough Building, Stanford
University, Stanford CA 94305-4045}

\begin{abstract}
 The single vortex problem in a strongly correlated bosonic system
is investigated self-consistently within the mean-field theory of the
Bose-Hubbard model. Near the superfluid-Mott transition, the vortex
core has a tendency toward the Mott-insulating phase, with the core
particle density  approaching the nearest commensurate value.
If the nearest neighbor repulsion exists, the charge
density wave order may  develop locally in the  core.
The evolution of the vortex configuration from the strong to weak coupling
regions is studied. This phenomenon can be observed in systems of
rotating ultra-cold atoms in optical lattices and Josephson junction
arrays.
\end{abstract}

\maketitle
Vortices in charged or neutral superfluids (SF) can be created
by applying an external magnetic field \cite{abrikosov1957} or by
rotating the system \cite{lifshitz1980}. Properties of vortices
are essential for understanding superfluidity
\cite{kleinert1989}.
Generally, vortices are  topological defects in the SF order
parameter with a $2\pi$ phase winding around the core. 
The SF order is suppressed near the core 
in an area characterized by  the size of the healing length $\xi$. 
As a result, other orders competing with the SF order may
develop there. In the context of high $T_c$ superconductors, it
was first proposed in the SO(5) theory that the vortex core is
antiferromagnetic in underdoped samples\cite{zhang1997}. This
theoretical prediction has now been confirmed by numerous
experiments\cite{levi2002}. Using the physics of the vortex core
to study competing orders of doped Mott insulators is an important
new direction in condensed matter physics\cite{sachdev2002}.

Quantized vortices have been observed in rotating systems
of dilute ultra-cold alkali atom gases
\cite{matthews1999,madison2000}, 
drawing intense attentions both experimentally and
theoretically \cite{anderson2000, madison2000a,chevy2000,
williams1999, fetter2001,baym1996,butts1999,baym2001}.
Most discussions up to now have been based on the
Gross-Pitaevskii-Bogoliubov (G-P-B) equations, which assume that
the particle density is given by the square of the amplitude of the SF 
order parameter. Thus, the minimum  particle density is always located 
at the core,  shown experimentally as a dark region in vortex imaging. 

Near the SF-Mott insulator (MI) transitions,
the G-P-B method ceases to work well.
Theoretically, many investigations based on the Bose-Hubbard model
are available now \cite{fisher1989,sheshadri1993} in the absence of rotation.
The MI phases occur at the commensurate (integer) fillings when the hopping 
amplitude $t$ is small compared with the Hubbard repulsion $U$. 
Transitions into the SF phase are triggered by either increasing $t$ or 
changing the chemical potential.  In the presence of the nearest neighbor 
repulsion $W$, the insulating CDW phases appear at half-integer fillings.
However, the vortex configuration in this region
has not been fully investigated.
Experimentally, tremendous progress has been made in realizing the SF-MI
transition in ultra-cold atomic systems in  optical lattices
\cite{greiner2002,orzel2001}, as proposed in Ref. \cite{jacksch1998}.
It would be interesting to further generate vortices 
in such systems.
Alternatively, Josephson junction  arrays and granular 
superconductors in magnetic 
fields are also possible systems to study such vortices, 
where Cooper pairs behave like composite  bosons. 

In this article, we address the vortex configuration near the SF-MI and 
SF-CDW transitions. In the former case, the vortex core is close to
the MI phase with a nearly commensurate filling while the SF
dominates in the bulk area. In the latter case, the superfluid
vortex with the CDW core behaves as a ``meron'' topological defect
of the 3-vector psuedospin order parameter.
These strong coupling configurations evolve to
the weak coupling G-P-B vortex smoothly as $t/U$ increases.
Although the discussion below is for the rotating neutral bosonic
system, it is also valid for the charged system in the magnetic field.
Theoretical predictions obtained in this paper can be tested in
these systems.

We study the 2D Bose-Hubbard model extended by a nearest neighbor
repulsion $W$ in the rotating  frame
\bea
H&=&-t \sum_{\langle \vec{r}_i, \vec{r}_j\rangle }
 \big \{ a^\dagger (\vec{r}_i) a( \vec{r}_j) e^{i\int_{r_j}^{r_i} d
\vec{r} \cdot \vec{A}(r)} +h.c. \big\}\nonumber \\
&+&\sum_{\vec{r}_i} \big \{ V_{ex} ( \vec{r}_i)-V_{cf}( \vec{r}_i)-\mu \big \}
 n(\vec{r}_i) \nonumber\\
&+& {U\over 2} \sum_i n(\vec{r}_i)  (n(\vec{r}_i)-1)
+W\sum_{\langle \vec{r}_i, \vec{r}_j\rangle } n( \vec{r}_i)  n( \vec{r}_j)
, \nonumber \\
\vec{A}(\vec{r}) &=& {m\over \hbar} \vec{\Omega} \times \vec{r}, ~~~~~~~~~~~
V_{cf}(\vec{r}_i)  =-{1\over 2} m \Omega^2 r_i^2,
\eea
with $\mu$ the chemical potential, $m$ the boson mass,
$\vec{\Omega}=\Omega \hat{z}$ the rotation angular velocity,
the position $\vec{r}$ related to the axis of rotation,
the vector potential ${\vec A}$ due to the Coriolis force,
$V_{cf}$ the centrifugal potential, and
$V_{ex}$ the trapping potential.
$U$ is scaled to 1.
For the charged bosonic system in the magnetic field $\vec{B}$,
$\vec{A}= (e^* / 2\hbar c)  \vec{B} \times \vec{r}$ instead and
the $V_{cf}$ term is absent.

\begin{figure}
\centering\epsfig{file=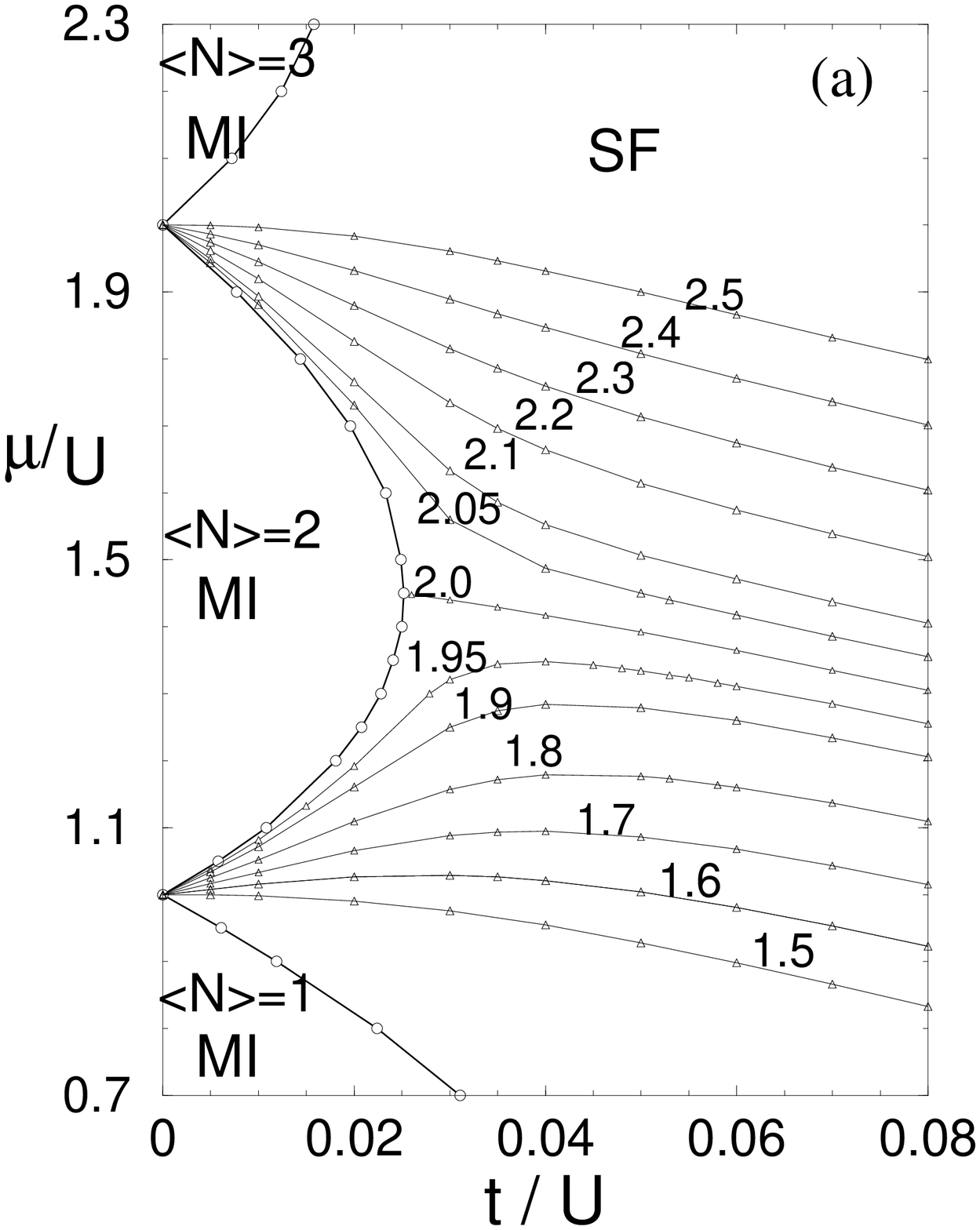,clip=1,width=42.5mm,angle=0}
\centering\epsfig{file=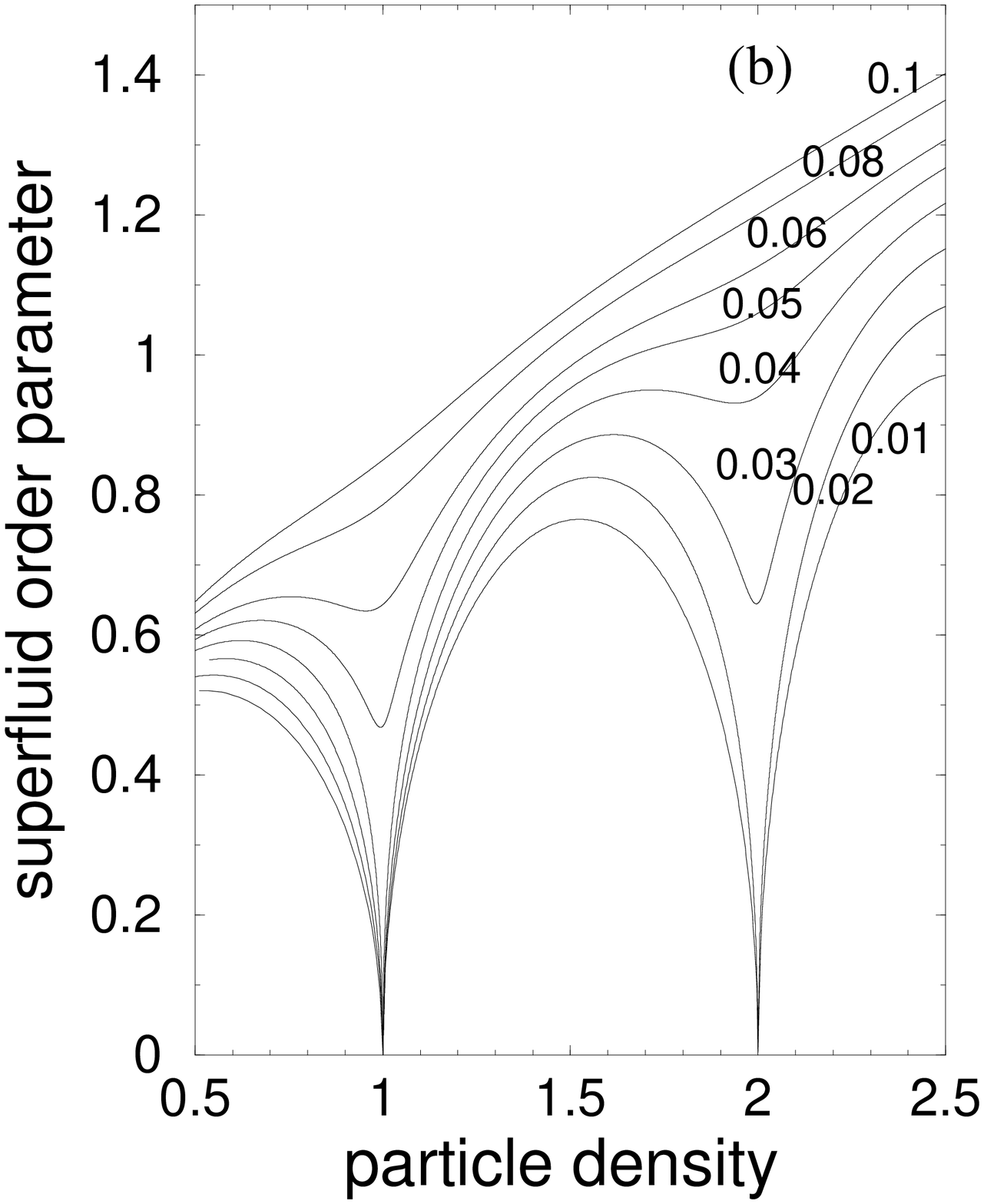,clip=1,width=43mm,angle=0}
\caption{(a) The ground state phase diagram for Bose-Hubbard model
at $W=0$ and $\Omega=0$. Lines of equal particle densities
are plotted with $\langle N\rangle=2.5\sim 1.5$ from top to bottom.
(b) The SF order parameter $\langle a\rangle$  versus
$\langle N \rangle$ at different values of $t/U=0.1\sim 0.01$
from top to bottom.
}\label{phase}
\end{figure}

\begin{figure}
\centering\epsfig{file=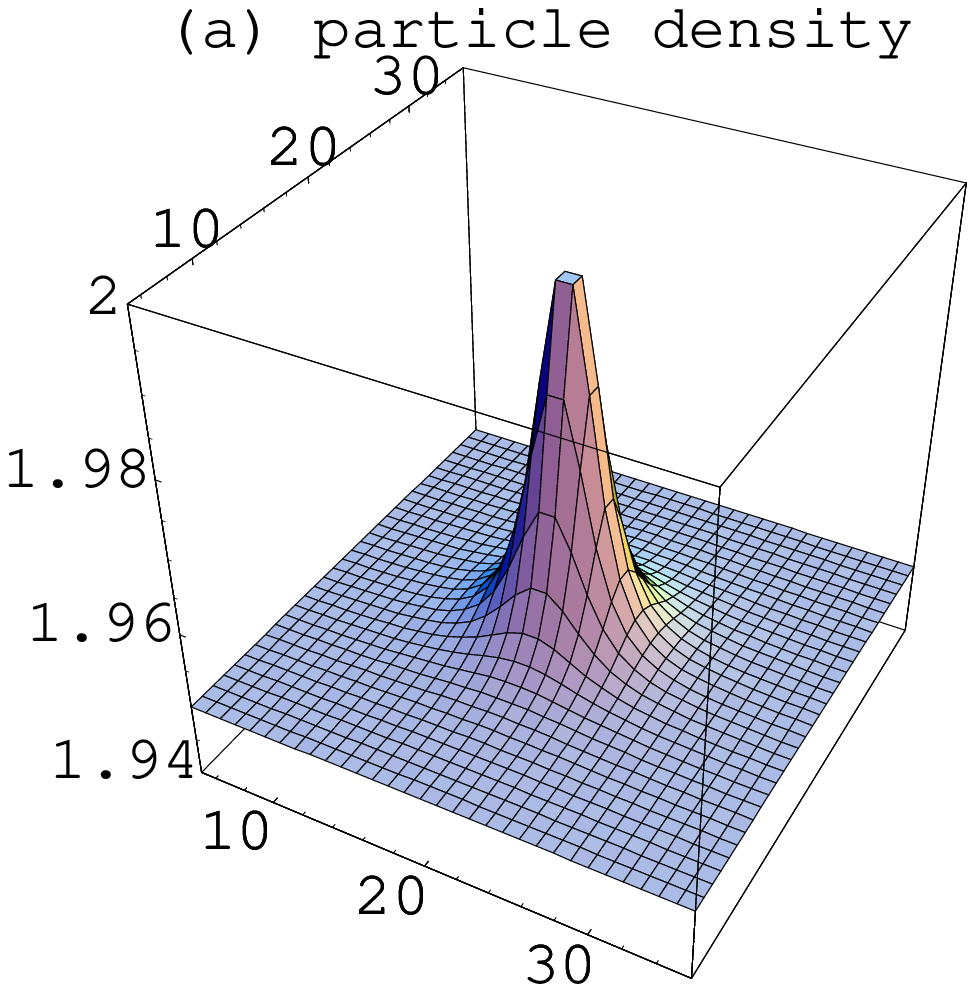,clip=1,width=42mm,angle=0}
\centering\epsfig{file=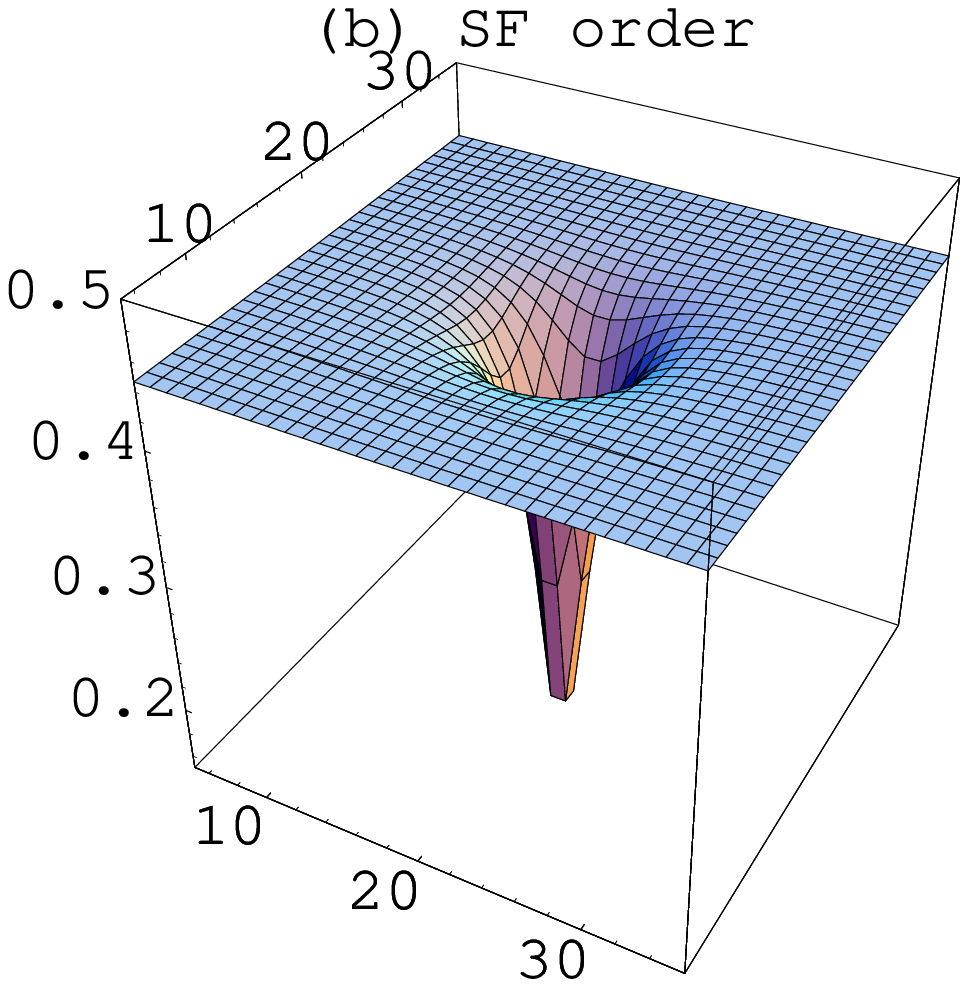,clip=1,width=42mm,angle=0}
\\
\centering\epsfig{file=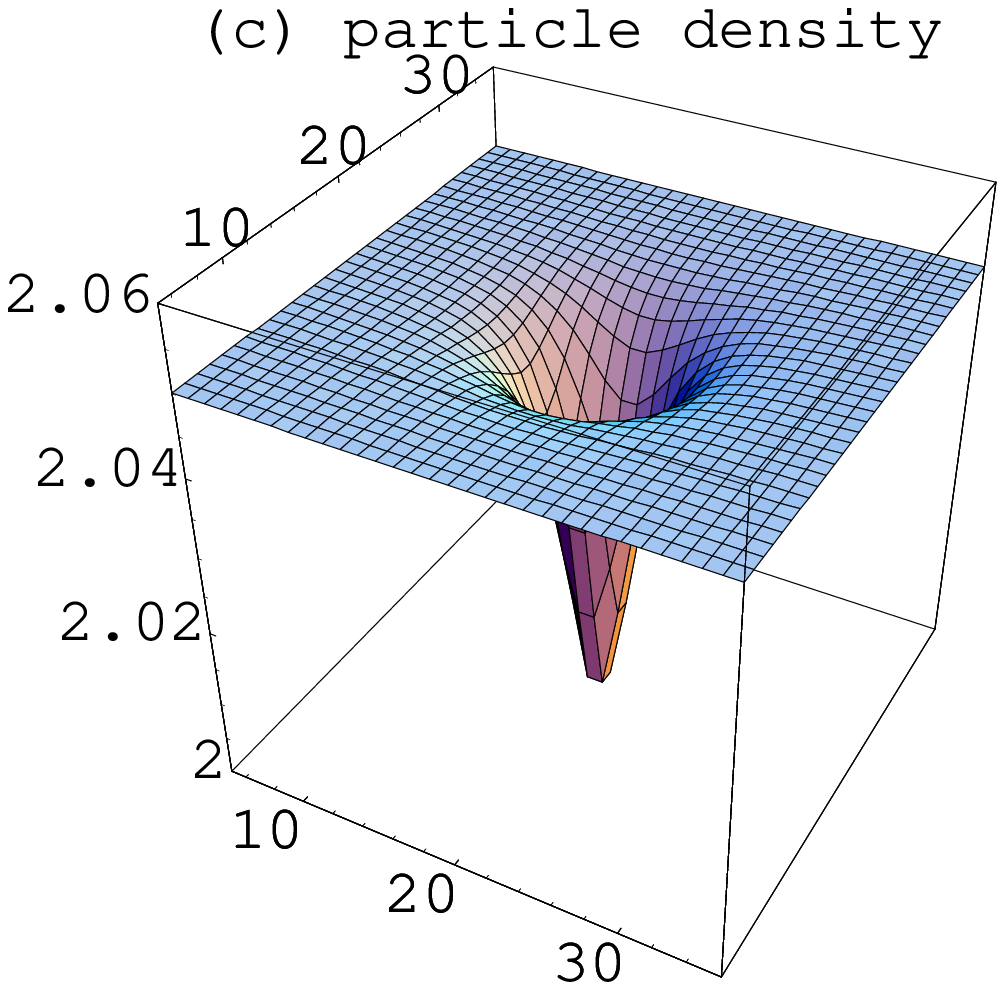,clip=1,width=42mm,angle=0}
\centering\epsfig{file=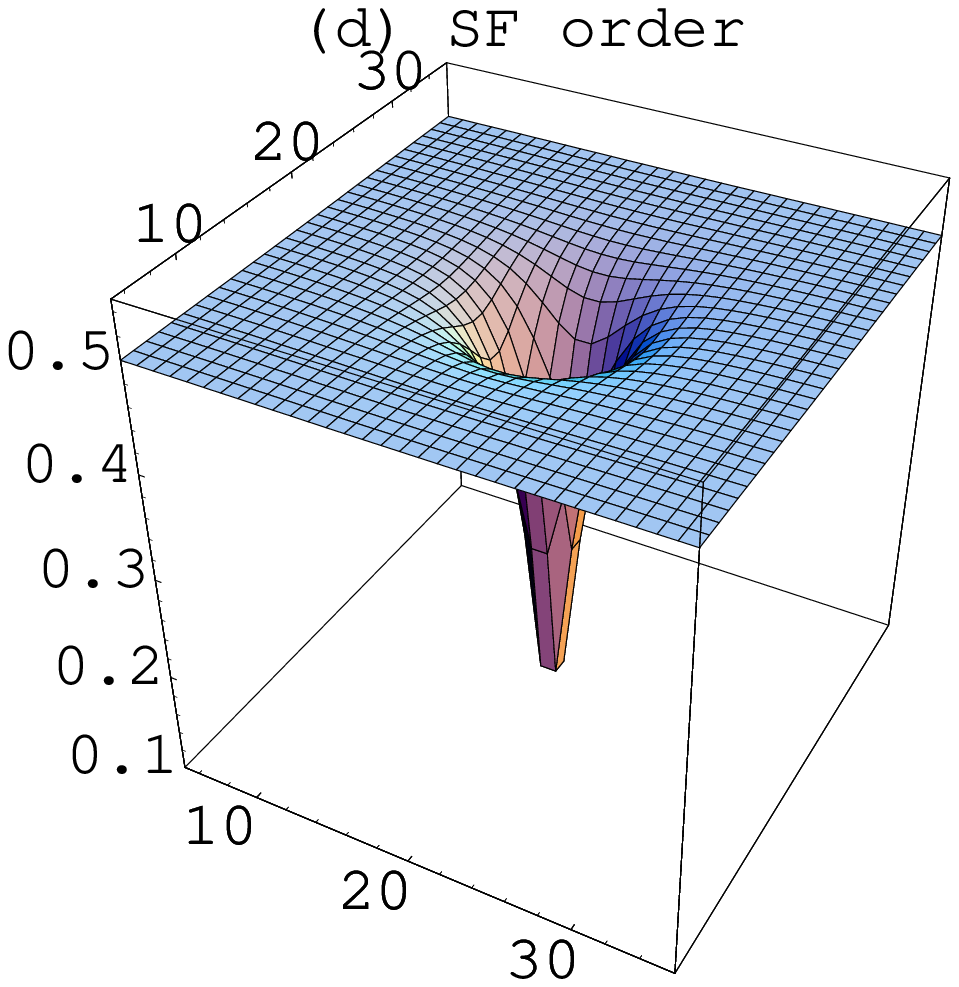,clip=1,width=42mm,angle=0}
\caption{ Vortex configurations near the SF-MI phase boundary at
$t/U=0.02$ with bulk particle density $\langle N \rangle=1.95$
(in (a) and (b) ) and $\langle N \rangle=2.05$ (in (c) and (d)).
The left hand side (a) and (c) are the particle density distributions;
the right hand side
(b) and (d) are the SF order distributions.
}\label{vort}
\end{figure}

This model can be solved self-consistently by the mean-field (MF)
approximation \cite{fisher1989,jacksch1998}, by decoupling the inter-site
terms as \bea a^\dagger(\vec{r}_i) a(\vec{r}_j) &\approx&
a^\dagger(\vec{r}_i) \langle a(\vec{r}_j)\rangle +
(i\leftrightarrow j) -\langle a^\dagger(\vec{r}_i) \rangle \langle
a(\vec{r}_j) \rangle,
\nonumber \\
n(\vec{r}_i) n(\vec{r}_j)& \approx & n(\vec{r}_i) \langle
n(\vec{r}_j) \rangle + (i\leftrightarrow j) - \langle
n(\vec{r}_i)\rangle \langle  n(\vec{r}_j)\rangle.\hspace{5mm} \eea
$\langle a(\vec{r}_j) \rangle$,  $\langle n(\vec{r}_j) \rangle$
are expectation values on the MF ground state, which is
approximated by a product wave function of the form
$|\Psi\rangle_{G}= \Pi_{\vec{r}_i} |\psi(\vec{r}_i) \rangle$. We
cutoff each single-site Hilbert space up to 10 particles which is
sufficient for experimental values $\langle N \rangle\approx 1\sim
3$ \cite{greiner2002}. The SF  and CDW order parameters  are defined
as $\langle a (\vec{r}_i) \rangle$ and $(-)^{r_i} (\langle n
(\vec{r}_i)\rangle - \langle N\rangle) $ respectively, where
$\langle N\rangle$ is the bulk average particle density. 
This simple MF theory describes the SF-MI transitions
and extrapolates well into the intermediate coupling region
\cite{fisher1989,sheshadri1993}.
For example,  the relation of the SF order $\avg{a}$ {\it v.s.} $t/U$
from the MF theory is in good agreement with that of the Monte-Carlo
simulations  at both commensurate (integer) and incommensurate
fillings\cite{krauth1991}.

\begin{figure}
\centering\epsfig{file=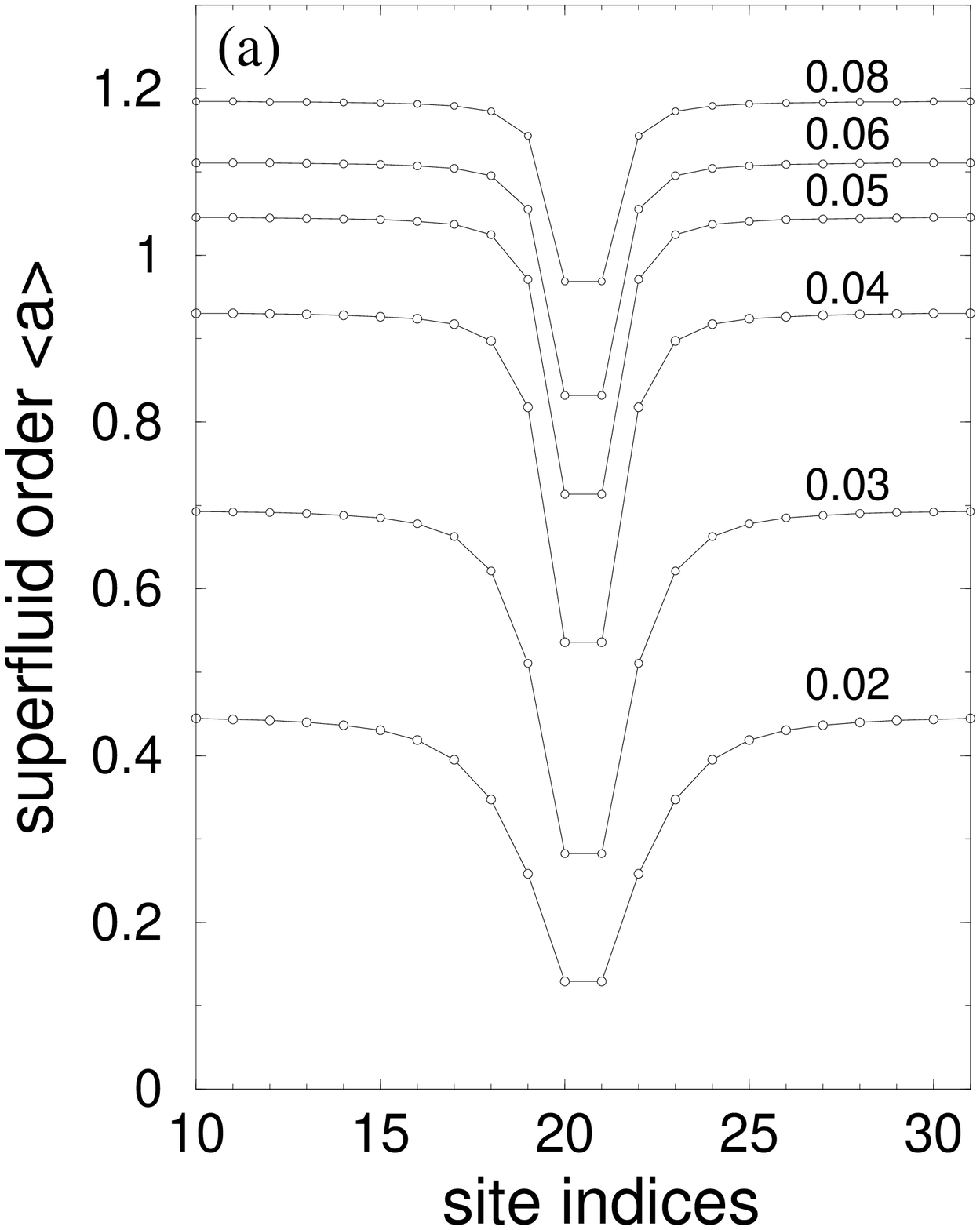,clip=1,width=42mm,angle=0}
\centering\epsfig{file=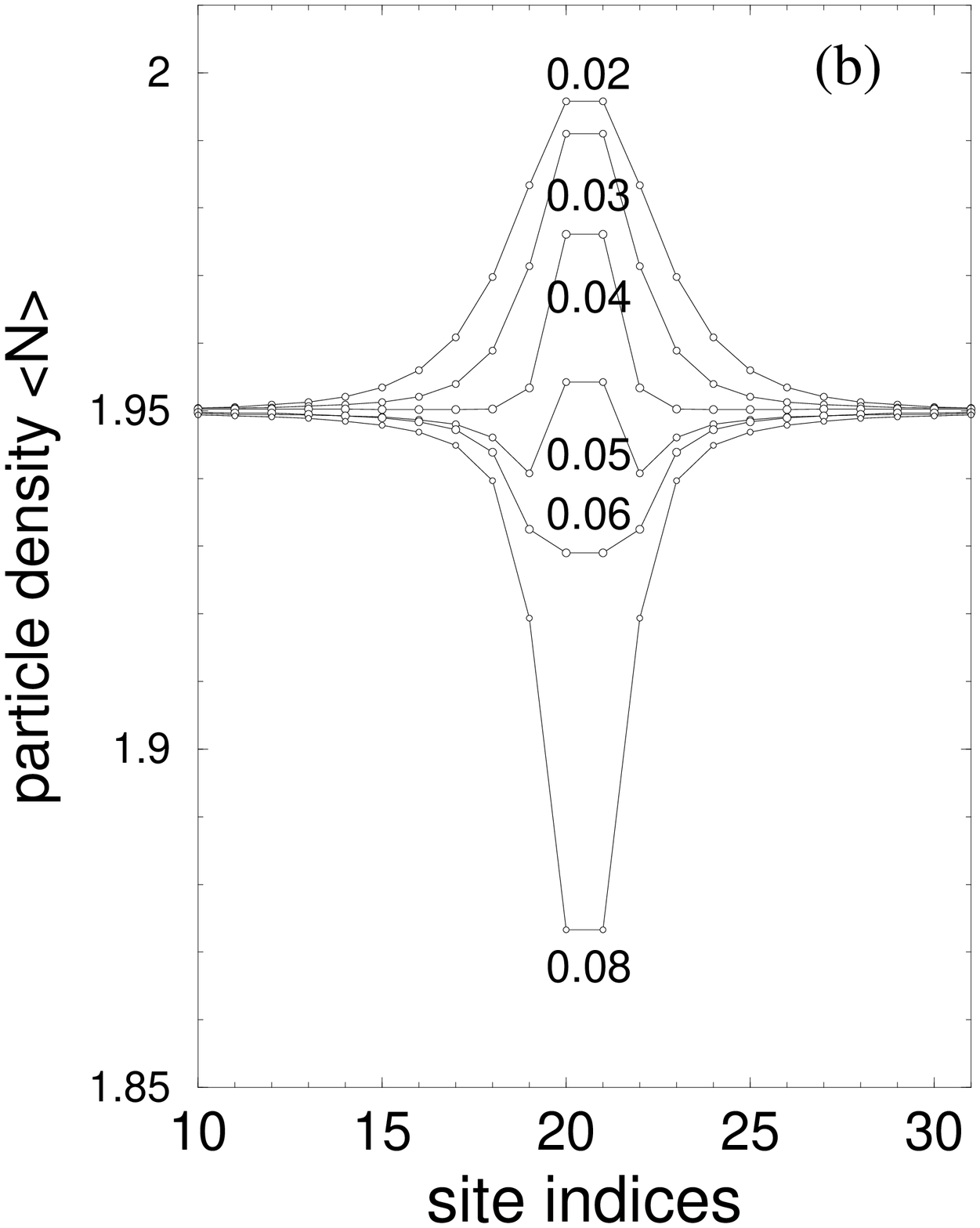,clip=1,width=42mm,angle=0}
\caption{The evolution of vortex configurations at $\langle N\rangle=1.95$
as increasing $t/U$.
SF amplitudes in (a) and particle densities in (b) are shown
along a path cut from (10,20) to (30,20) in the $40 \times 40 $ system.
$t/U=0.02\sim0.08$ from bottom to top in (a) and top to
to bottom in (b).
}\label{evolve}
\end{figure}

Before discussing the vortex problem, it is helpful to 
recall the well-known phase diagram when $\Omega=0$
\cite{fisher1989} (Fig. \ref{phase} (a)).
From the  equal density contours we can see an
approximate particle-hole (p-h) symmetry with
respect to $\langle N \rangle= 2$ near the phase boundary:
$\mu$ drops or increases with increasing $t/U$ when $\langle N \rangle\ge
2$ or $<2$ respectively.
These can be viewed as particle or hole SF,
but this difference no longer exists as $t/U$ becomes large,
where $\mu$ drops with increasing $t/U$ for both cases, i.e.,
the system evolves from the strong to weak coupling region.
In Fig. \ref{phase} (b), SF order  ${\it vs.}$  $\langle N \rangle$
is shown at different values of  $t/U$.
The SF order parameter increases monotonically with $\avg{N}$
at large $t/U$, while the commensurate filling suppresses the SF
order prominently at small $t/U$, eventually leading to the MI phase.
At $t/U\approx 0.08$, the suppression disappears
even at $\langle N \rangle=1$, which marks the cross-over into the
weak coupling region.

\begin{figure}
\centering\epsfig{file=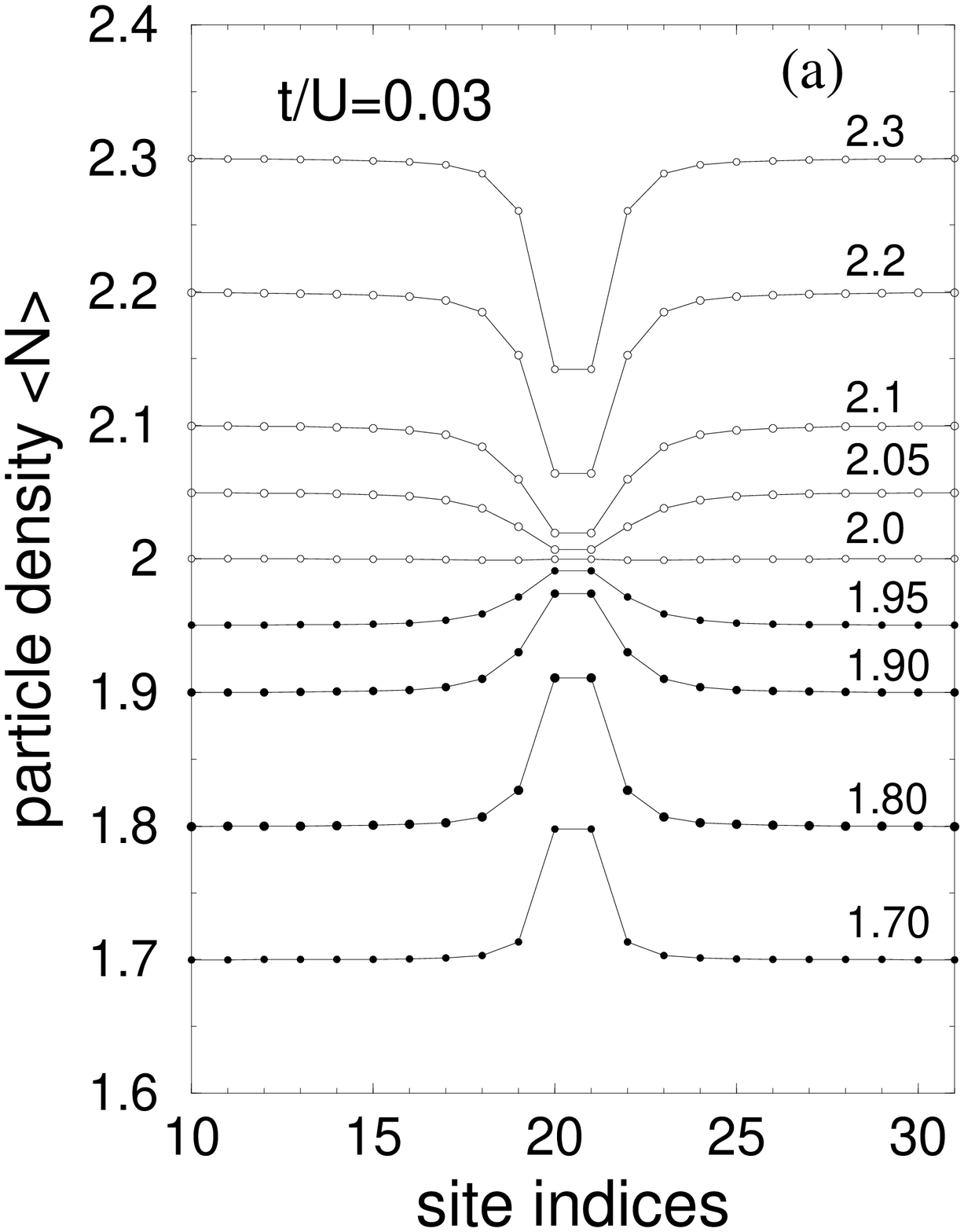,clip=1,width=42mm,angle=0}
\centering\epsfig{file=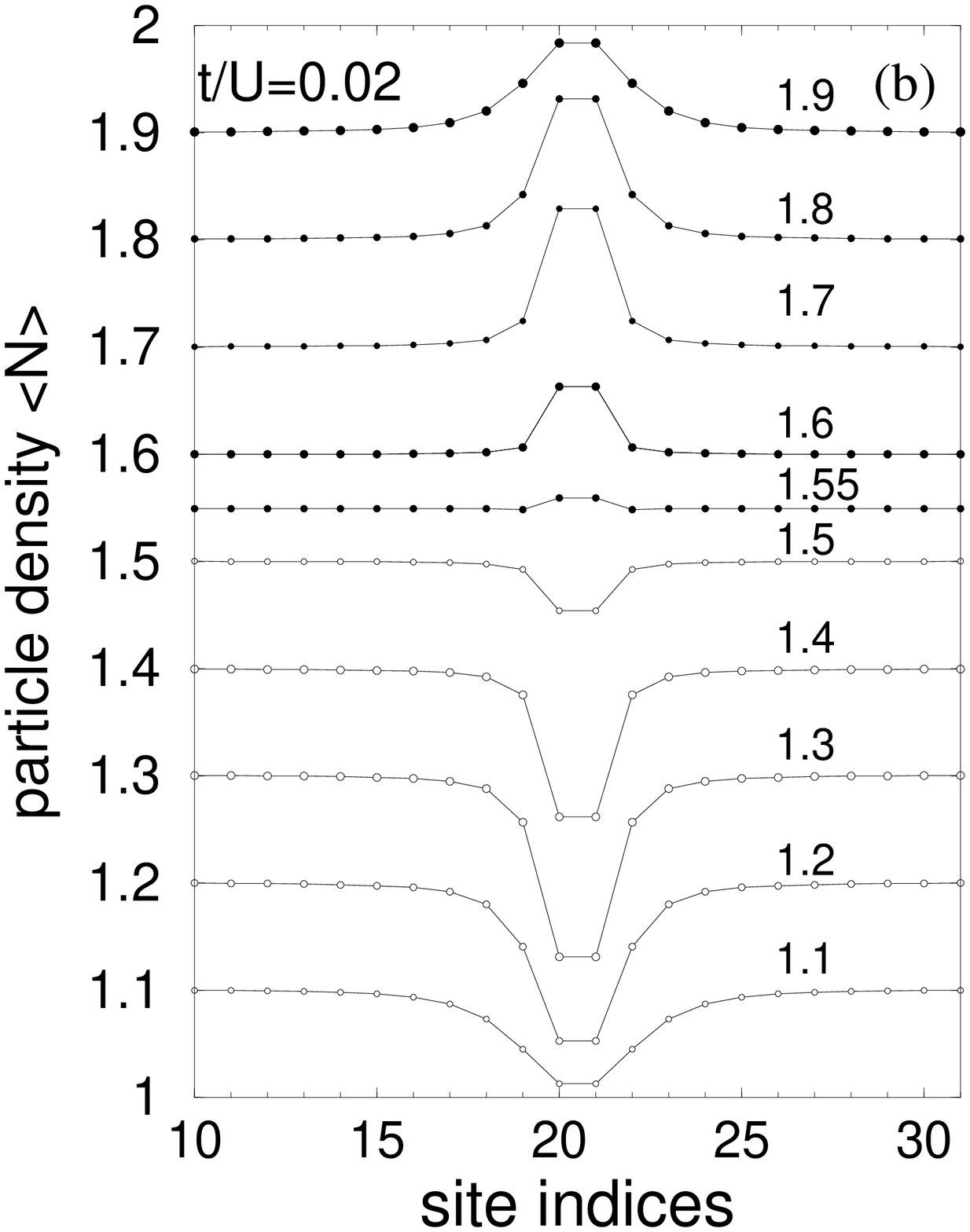,clip=1,width=42mm,angle=0}
\caption{ The evolution of vortex particle density distribution
at fixed $t/U=0.03~(a)$ and $0.02~(b) $  with varying bulk values
$\langle N \rangle$ along the same path in Fig. \ref{evolve}.
(a) From top to bottom,  $\langle N \rangle=2.3\sim 1.7$.
(b) From top to bottom,  $\langle N \rangle=1.9 \sim 1.1$.
}\label{fixt}
\end{figure}

\begin{figure}
\centering\epsfig{file=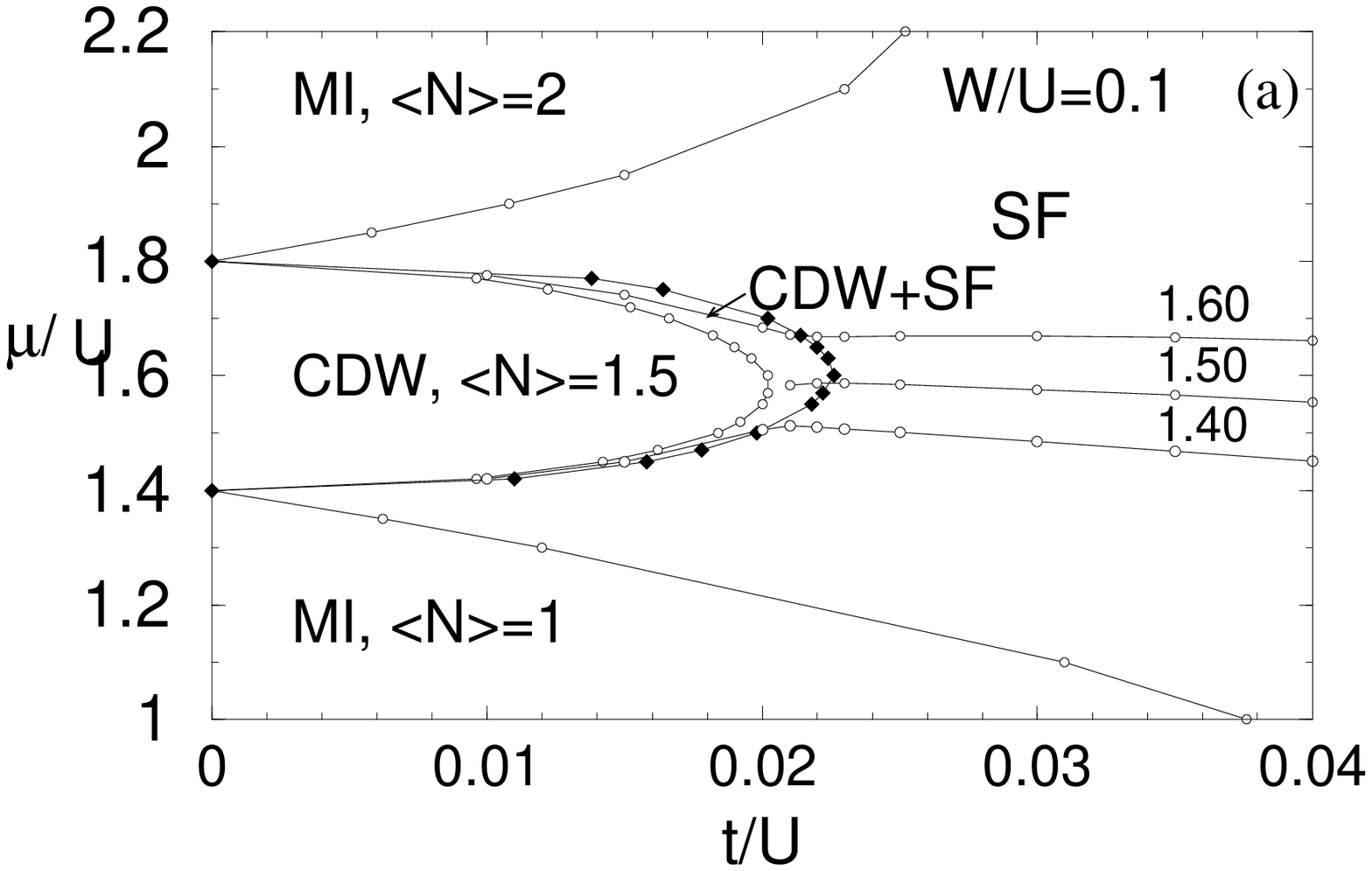,clip=1,width=8cm,angle=0} 
\\
\centering\epsfig{file=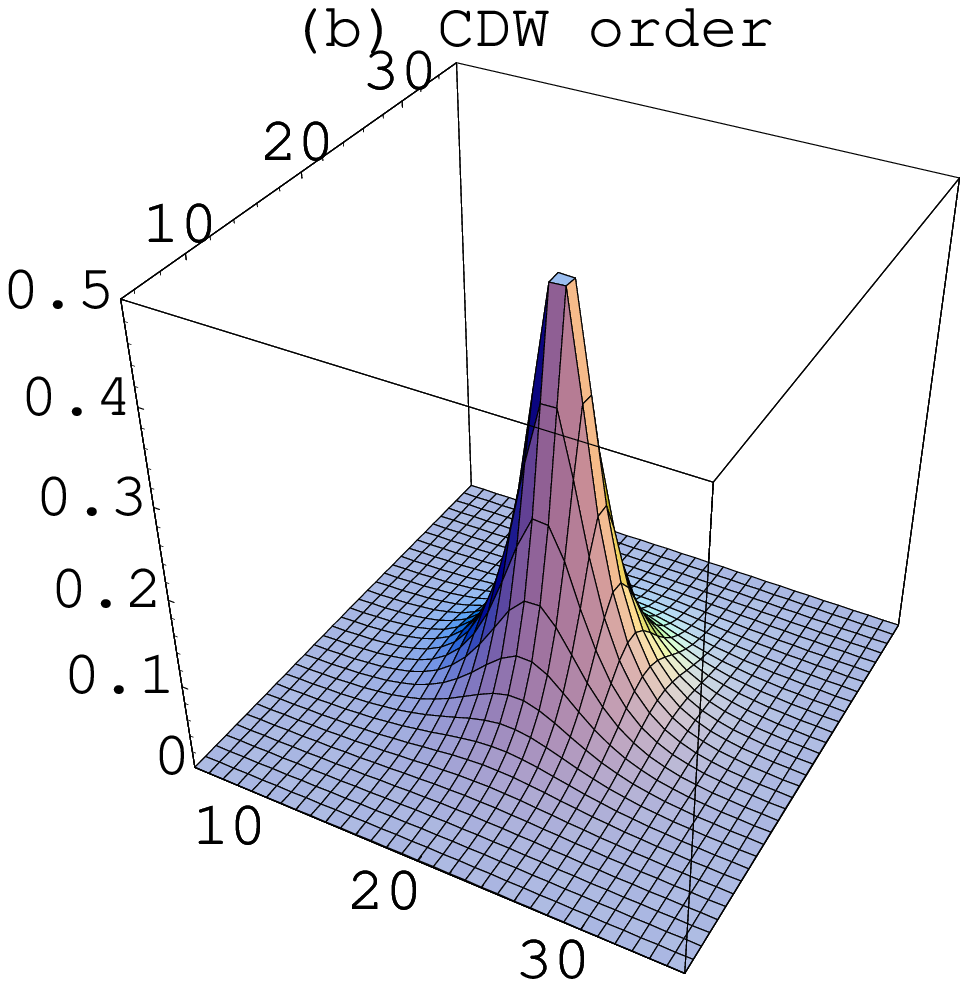,clip=1,width=42.5mm,angle=0}
\centering\epsfig{file=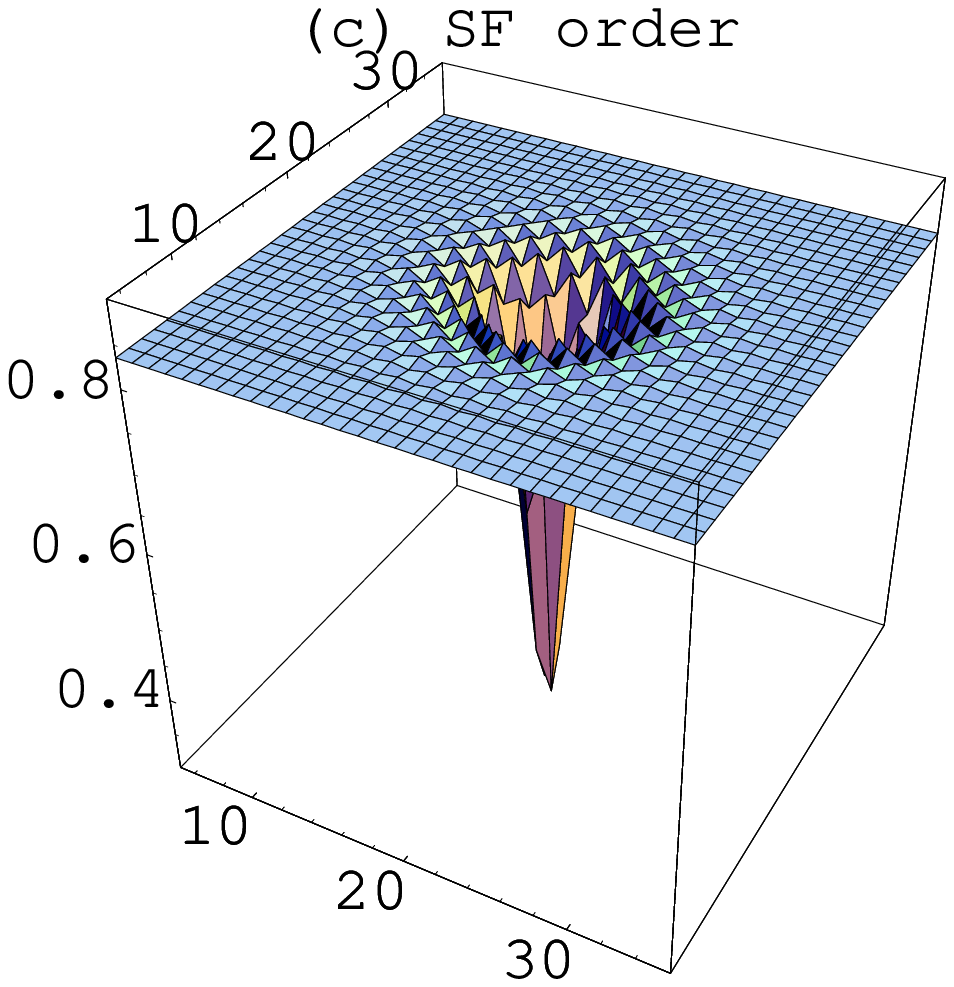,clip=1,width=42.5mm,angle=0}
\caption{(a)
Phase diagram of the extended Bose-Hubbard model with $W/U=0.1$
around $\langle N \rangle =1.5$.
Lines of equal particle densities are plotted with
$\langle N \rangle=1.6\sim 1.4$ from top to bottom.
The CDW (b) and SF (c) order distributions
in the  vortex configuration near the CDW phase
with $t/U=0.023$ and bulk average $\langle N \rangle =1.5$.
}\label{cdw}
\end{figure}

We study the single vortex problem in a 
$40 a_0 \times 40 a_0$ ($a_0$ being the lattice constant)
system around which the circulation of $\vec{A}$  is $2\pi$ and thus $\Omega=
h/(2mL^2) (L=40 a_0)$ correspondingly.
The rotation center is located at the center of the central plaquette.
We further simplify the problem by dropping the $V_{ex}$ and $V_{cf}$ terms,
since they behave smoothly near the center
of the trap and they can cancel each other if the $\Omega$ is close
to the trap frequency.

Two typical vortex configurations near the SF-MI transition are
shown in Fig. \ref{vort}  (a) $\sim$ (d) with $t/U=0.02$,
where the particle densities in  cores are maximal or minimal respectively.
The vortex core is located at the center of the plaquette with the
reduced SF order on the sites nearby.
Roughly speaking, the square of the superfluid amplitude,
$|\langle a \rangle|^2$, is proportional to  $|\langle N \rangle -N_0|$
near the transition, where $N_0=2$ here is the nearest commensurate value.
The local particle density in the core should be closer to the commensurate
value in order to suppress the SF order.
As a result, the particle density reaches a maximum or minimum
when the bulk density $\langle N \rangle$ is slightly smaller or larger
than the commensurate value.
The former case can also be understood as the vortex of the hole superfluid,
where the hole density goes to a minimum at the core.
This contrasts with the case of  fermionic superfluidity,
where the Cooper pairs are broken into normal particles in the core
with total particle density almost unchanged,
and also with the case of the weak coupling bosonic system where the only
possibility is the depletion of the core particle density.
As we approach the vortex core from outside, the hopping process is frustrated
and thus effectively $t/U$ becomes small because of the phase winding
of the SF order.
As a result, the vortex core is driven closer to the MI state
than the bulk area.
We also check the vortex configuration with integer value of $\avg{N}=2$
at the same value of $t/U$, where the particle density distribution is
uniform with suppressed SF order in the core.

We further discuss the evolution of the vortex core configuration
from the strong to the weak coupling regions.
The bulk particle density is fixed at $\langle N \rangle=1.95$ with
increasing $t/U$ as shown in Fig. \ref{evolve}.
The SF order increases and the healing length $\xi$ decreases
away from the SF-MI boundary along this direction of evolution.
On the other hand, the weak coupling expression $\xi/a_0 = \sqrt{t/(U
\langle N \rangle) }$
states that $\xi$ increases with increasing $t/U$.
Thus we can infer that $\xi$'s minimum value appears in the intermediate
coupling region.
The vortex core with extra particle density survives until
$t/U\approx 0.05\sim 0.06$, after which it crosses over gradually to
that with depleted particle density at large $t/U$.
This transition agrees with the behavior of the SF order
{\it vs.} $\avg{N}$ when the system is not in rotation.
In that case, at the same filling level, we check that the suppression
due to the commensurate filling also vanishes around a similar value
of $t/U$.
When $t/U$ is larger than $0.06$, the vortex configuration is already
similar with the weak coupling case.
We also check the above evolution with fixed  $\langle N\rangle=2.05$.
The feature of SF order is  similar with that in Fig. \ref{evolve} (a),
and the minimum particle density is always located in the core.
When $t/U$ is less than an intermediated value around 0.05,
the core particle density is close to  the commensurate value $2.0$.
When $t/U$ grows larger, it  drops further.
This also agrees with the evolution picture
from  strong to weak coupling physics.

Another evolution from the particle-like vortex core to the hole-like core
with fixed $t/U$ in the strong coupling region and  varying
$\langle N \rangle$ is shown in Fig. \ref{fixt} (a).
We choose the region close to the MI phase with $\avg{N}=2$, where the
approximate p-h symmetry is valid, with $t/U$ fixed at 0.03 and
$\langle N \rangle$ varying from 2.3 to 1.7.
As a result, the difference between the core particle density
and the bulk value changes from negative to positive
as the bulk density passes $\langle N \rangle=2$.
This point is the closest one to the tip of the MI phase along the evolution,
where  the minimum of the SF order and the maximum of $\xi$ also lie.
In Fig. \ref{fixt} (b), such evolution is shown along the path 
$t/U=0.02$ connecting two neighboring MI phases with
$\langle N \rangle=2$ and  1.
The core configuration is close to the MI phase with
$\langle N \rangle=2$ on the top, and becomes close to the MI phase
with $\langle N \rangle=1$ at the bottom.
Around $\langle N \rangle=1.55$, the density distribution becomes
almost uniform.
This point is the  maximum of the SF order
and the minimum of $\xi$ because it is the farthest point from
the MI, which is just opposite to Fig. \ref{fixt} (a).
In both Fig. \ref{fixt}(a) and (b), the tendencies to MI phase
in the vortex core are strong when the average $\avg{N}$ is
close to an integer, and becomes weaker as $\avg{N}$ away from
commensurate fillings.

Next we turn on the nearest neighbor repulsion $W/U=0.1$.
In the absence of rotation, CDW phases appear between two neighboring
commensurate MI phases at half-integer fillings when $t/U$ is small
\cite{pich1998}.
At large values of $t/U$, SF phases stabilize as usual.
Between the CDW and the SF phases, the mean field theory gives a small area
of the coexistence of CDW and SF order, i.e., the supersolid phase
\cite{pich1998}.
These are shown in Fig. \ref{cdw} (a) with the equal density lines
around $\langle N \rangle=1.5$.
It is well known that the hard-core boson model can be mapped into spin
1/2 XXZ model in a magnetic field, and thus the CDW and SF orders can
be unified in a 3-vector pseudospin picture.
With the releasing of the hard core constraint, the above mapping
is still approximately valid in the sense that the spin up and down states
correspond to two nearest integer number states on each site.
The vortex configuration at $t/U=0.023$ with  bulk particle density
$\langle N \rangle=1.5$ is shown in Fig. \ref{cdw} (b) and (c).
The SF order dominates  outside the core, while the CDW order develops
together with the suppression of the SF order in the core.
In the pseudo-spin picture, this is a kind of topological defect
called ``meron'', where the pseudo-spin pointing along z axis at the origin
gradually changes to lying in the x-y plane with a winding number 1
around the origin when far from it.
This vortex configuration also evolves to the weak coupling one
at large value of $t/U$ with the disappearance of the CDW order in the core.
This situation is similar to the behavior of the superspin
in the $SO(5)$ theory of the antiferromagnetic vortex 
core\cite{zhang1997,levi2002}.

In previous experiments, the vortex core with size $2\xi\approx0.4\mu$m is too 
small to observe directly by optical methods.
A time-of-flight expansion is needed before optical absorption 
imaging\cite{madison2000}.
On the optical lattice, the vortex core is larger.
The typical core size in our calculations
is estimated at $ 5\sim 6$ lattice constant $a_0$ ($a_0=0.426~ \mu$m in 
Ref. \cite{greiner2002}), i.e. about $2 ~\mu$m. 
The resonant probe laser beam can be focused to this size  
at the level of  current technology \cite{chu}.
Thus without turning off the trap, it is possible to image the core particle 
density distribution non-destructively by scanning the probe laser beam.
It would be interesting to find the anomalous vortex with the maximum
particle density in the core.
Another possible realization is the Josephson junction array.
The non-uniform charge distributions in the vortex configuration result 
in electric fields \cite{beasley}.
Thus it is possible to determine the filling in the vortex core with respect
to the outside by measuring electric field distributions.

In summary, we have studied vortex structures of the strong coupling
boson systems.
Near the SF-MI transition, the vortex core is  more strongly coupled
compared to the bulk area and is thus closer to the MI phase with
suppressed SF order.
The particle density in the core can be either the maximum or the minimum
of the whole system, always approaching the nearest commensurate density
of the Mott insulator.
Near the SF-CDW transition, a superfluid meron-like vortex is found with
a CDW core.
All of  these strong coupling vortex configurations evolve to the
conventional weak coupling one as $t/U$ increases.

We thank  G. Baym,  M. Beasley, B. A. Bernevig,
I. Bloch,  C. Chin,  S. Chu,  A. L. Fetter,  N. Gemelke,
P. SanGiorgio, E. Mukamel, and F. Zhou  for helpful discussions.
This work is supported by the NSF under grant numbers DMR-9814289, and
the US Department of Energy, Office of Basic Energy Sciences under
contract DE-AC03-76SF00515. CW and HDC are also supported by the
Stanford Graduate Fellowship program.


\begin{thebibliography}{10}
\bibitem{abrikosov1957}
A. A. Abrikosov, Soviet Phys. JETP, {\bf 5} 1174(1957).

\bibitem{lifshitz1980}
E. M. Lifshitz and L. P. Pitaevskii, {\it Statistical Physics} part 2,
3rd edn(Oxford:Pergamon) (1980).

\bibitem{kleinert1989}
H. Kleinert, Gauge fields in condensed matter, Vol I
World Scientific,  Singapore (1989);
H. Kleinert, Lett.\ Nuovo Cimento  {\bf 35}, 405 (1982).

\bibitem{zhang1997}
S.C.Zhang, Science, 275, 1089 (1997);
D. Arovas {\it et al}, \PRL {\bf 79},
2871(1997),

\bibitem{levi2002}
B. G. Levi, Physics Today, {\bf 55 }(2), 14-16 (2002).

\bibitem{sachdev2002}
S. Sachdev, S.C. Zhang, Science {\bf 295}, 452 (2002).

\bibitem{matthews1999}
M. R. Matthews {\it et al}, \PRL ~{\bf 83}, 2498(1999).


\bibitem{madison2000}
K. W. Madison {\it et al}, \PRL ~{\bf 84}, 806(2000).

\bibitem{anderson2000}
B. P. Anderson {\it et al}, \PRL ~{\bf 85}, 2857 (2000).

\bibitem{madison2000a}
K. W. Madison {\it et al}, J. Mod. Opt. {\bf 47}, 2715(2000).

\bibitem{chevy2000}
F. Chevy {\it et al}, \PRL ~{\bf 85}, 2223(2000).

\bibitem{williams1999}
J. E. Williams {\it et al} Nature 401, 568 - 572 (1999).

\bibitem{fetter2001}
A. L. Fetter {\it et al},
J. Phys. : Condens. Matter {\bf 13}, R135-R194 (2001).

\bibitem {baym1996}
G. Baym {\it et al}, \PRL~ {\bf 76}, 6 (1996).



\bibitem{butts1999}
D. A. Butts{\it et al}, Nature, {\bf 397}, 327(1999).

\bibitem{baym2001}
U. R. Fischer and G. Baym, \PRL~ {\bf 90}, 140402 (2003).


\bibitem{fisher1989}
M. P. A. Fisher {\it et al} \PRB~{\bf 40}, 546 (1989).


\bibitem{sheshadri1993}
K. Sheshadri {\it et al}, Europhys. Lett. {\bf 26}, 545(1993).

\bibitem{greiner2002}
M. Greiner, {\it et al}, Nature {\bf 415}, 39(2002).

\bibitem{orzel2001}
C. Orzel,  {\it et al}, Science {\bf 291}, 2386(2001).


\bibitem{jacksch1998}
D. Jaksch, C. Bruder, J. I. Cirac, C. W. Gardiner, and P. Zoller,
\PRL {\bf 81}, 3108(1998).



\bibitem{krauth1991} W. Krauth {\it et al}, \PRL ~{\bf 67}, 2307(1991).

\bibitem{pich1998} C. Pich {\it et al}, \PRB ~{\bf 57}, 13712(1998).

\bibitem{chu}
S. Chu, private communication.

\bibitem{beasley}
M. Beasley, private communication.

\end{thebibliography}
\end{document}